# Manipulating freely diffusing single 20-nm particles in an Anti-Brownian Electrokinetic Trap (ABELtrap)

Nawid Zarrabi[a], Caterina Clausen[a], Monika G. Düser[a], Michael Börsch[a,b,]*

[a] 3rd Institute of Physics, University of Stuttgart, Pfaffenwaldring 57, 70550 Stuttgart, Germany
[b] Single-Molecule Microscopy Group, Jena University Hospital, Friedrich Schiller University Jena,
Nonnenplan 2 - 4, 07743 Jena, Germany

**ABSTRACT**

Conformational changes of individual fluorescently labeled proteins can be followed in solution using a confocal microscope. Two fluorophores attached to selected domains of the protein report fluctuating conformations. Based on Förster resonance energy transfer (FRET) between these fluorophores on a single protein, sequential distance changes between the dyes provide the real time trajectories of protein conformations. However, observation times are limited for freely diffusing biomolecules by Brownian motion through the confocal detection volume. A. E. Cohen and W. E. Moerner have invented and built microfluidic devices with 4 electrodes for an Anti-Brownian Electrokinetic Trap (ABELtrap). Here we present an ABELtrap based on a laser focus pattern generated by a pair of acousto-optical beam deflectors and controlled by a programmable FPGA chip. Fluorescent 20-nm beads in solution were used to mimic freely diffusing large proteins like solubilized $F_oF_1$-ATP synthase. The ABELtrap could hold these nanobeads for about 10 seconds at the given position. Thereby, observation times of a single particle were increased by a factor of 1000.

## 1 INTRODUCTION

The mechanochemistry of the membrane enzyme $F_oF_1$-ATP synthase has been the main focus of our research in the last 15 years[1-30]. The enzyme provides the fundamental 'chemical energy currency' adenosine triphosphate (ATP) for all kinds of living cells. It is located in the inner mitochondrial membrane, in the thylakoid membrane and in plasma membranes of bacteria. Catalysis is driven by mechanochemical coupling of rotary subunit movements within the enzyme which induce and synchronize conformational changes in the three ATP binding sites. The rotation of subunits in $F_oF_1$-ATP synthase can be summarized briefly. Proton translocation through the membrane-bound $F_o$ motor of ATP synthase from *Escherichia coli* powers a 10-step rotary motion of the ring of *c* subunits[13]. This rotation is transmitted to the γ and ε subunits of the $F_1$ motor, which rotate in 120° steps at high driving forces[4, 6]. We aim to unravel the movements of the two motors in real time by monitoring subunit rotation using single-molecule Förster resonance energy transfer (FRET). Therefore one fluorophore can be attached specifically to the $F_1$ motor, another one to the $F_o$ motor of the liposome-reconstituted enzyme[2]. During rotation the distance between the two markers is changing stepwise. Time trajectories of FRET changes of single enzymes unraveled many details of the conformational dynamics. Photophysical artifacts due to spectral fluctuations of the single fluorophores were omitted from conformational analysis by a previously developed duty cycle-optimized alternating laser scheme (DCO-ALEX)[9]. Recently, using three fluorophores attached to a single $F_oF_1$-ATP synthase allowed to simultaneously monitor both rotary motors with their distinct step sizes[15].

The main drawback of our confocal single-molecule FRET approach using freely diffusing, liposome-reconstituted FoF1-ATP synthase is the limited observation time. Stochastic Brownian motion of the proteoliposomes through the confocal excitation and detection volume yields time trajectories with a maximum length in the range of several hundreds milliseconds. The average diffusion time of our 120-nm proteoliposomes is about 30 ms. The dwell times for the individual rotary motor steps were found in the range of 6 ms to 25 ms at high ATP concentrations for ATP hydrolysis or at a high proton motive force for ATP synthesis, respectively. Therefore longer observation times are required to investigate the rotary motions at lower driving forces or in the presence of inhibitors.

..................................................................................

email: michael.boersch@med.uni-jena.de or m.boersch@physik.uni-stuttgart.de ; http://www.m-boersch.org

Our attempts to immobilize the proteoliposomes via biotin lipids to a streptavidin-coated cover glass were not successful, because we could identify the expected FRET levels in FoF1-ATP synthase but failed to observe the ATP-driven FRET changes by subunit rotation[9]. Thus we were seeking for alternative methods to hold the FRET-labeled enzyme in place in solution, i.e. within the confocal volume.

In 2005 we met A. E. Cohen and W. E. Moerner who presented their microfluidic device to hold single small particles like fluorescent beads, liposomes, DNA or proteins in solution[31-33]. The so-called ABELtrap (Anti-Brownian Electrokinetic Trap") comprises a 1-µm thin, crossed channels structure with four access channels. In each access channel one platinum electrode is placed. The fluorescence of the particle is used to localize its position in x and y coordinates. Depending on the actual distance to a chosen target position within the trapping region, voltages are supplied to the electrodes and the particle is moved by electrophoretic and electroosmotic forces. Thereby the arbitrary Brownian motion of the particle is fully compensated and the particle remains stationary. Meanwhile several ABELtrap versions have been realized[34-38], with very fast feedback times in the microsecond time range[39-43], and similar devices have been published by other researchers[44, 45].

We started with an ABELtrap approach at the University of Stuttgart using an EMCCD camera for particle localization[46] according to the published setup at Stanford University. Adding 10 to 50 NIR-fluorescently labeled lipids to the single liposome should be used to localize the liposome and to push it back to a target position. NIR dyes were excited with a 785-nm laser diode and fluorescence was detected around 800 nm. At the target position, we placed the confocal excitation volume of a second laser (488 nm) to measure FRET of the reconstituted single FoF1-ATP synthase in this liposome. Liposomes could be trapped for several seconds. The proof of principle has been shown using liposomes containing two different fluorescent lipids, and details of this setup were published[22, 23, 46]. However, the photostability of the NIR dye was poor and the number of labeled lipids per liposome could not be controlled. Apparently, the presence of the NIR dyes affected the brightness of the single fluorophores on the enzyme, eventually by additional FRET from the fluorophores on the protein to the labeled lipids. In addition, the position of the trapped liposome was still fluctuating, and the related fluctuating fluorescence intensities of the FRET-labeled enzyme did not yield a stable FRET time trajectory.

Here we report the setup of a faster, confocal ABELtrap built according to the published microscopy systems by A. E. Cohen and W. E. Moerner. The confocal ABELtrap can use the intensity of a single fluorophore to localize the particle, and the feedback is fast enough to even trap a single soluble fluorophore in solution. We modified the public available software[40] to control the confocal laser focus using a pair of acousto-optical beam deflectors. With this setup, trapping of 20-nm fluorescent beads for 10 seconds was possible, and further improvements will allow holding single FRET-labeled FoF1-ATP synthases in liposomes in buffer solution.

## 2 EXPERIMENTAL PROCEDURES

### 2.1 Setup of the confocal ABELtrap with AOBDs

In the confocal ABELtrap, the laser focus is moved in a 2 µm x 2 µm area within the 1-µm deep trapping region. A pattern of laser positions is generated. Once a fluorescent particle is excited, the position of the fluorescence intensity is estimated and a feedback voltage is created to move the particle into the center of the laser pattern. We use a pair of AOBDs to shift the laser focus in x and y coordinates. A sketch of the optical pathway is shown in Figure 1a, and the image of the setup in Figure 1b.

For excitation with 491 nm we used a solid-state continuous-wave laser (Cobolt Calypso, 50 mW). The laser beam was focused by lens L1 (200 mm) into a position between the two crystals of the AOBDs (Gooch & Housego, NEOS technologies, 46080-3-LTD) which were aligned and mounted on a mechanically adjustable stage (mounting platform 71001). The diffracted first order beam was selected by the subsequent aperture B1 and divergence was corrected with the second lens L2 (100 mm). A pair of lenses L3 (25 mm) and L4 (200 mm) expanded the beam diameter to illuminate the back aperture of the microscope objective O1 (PlanApo 100x oil, N.A. 1.35, Olympus). A dichroic beam splitter BS1 (dual line z 488 / 561, AHF Tübingen, Germany) rejected scattered laser light from fluorescence photons in the detection pathway. An achromatic lens L5 (200 mm) imaged the confocal excitation volume onto the center of the pinhole P1 (150 µm), which was mapped again to the detection area of the APDs by lenses L6 or L7 (50 mm). Fluorescence was separated into two channels for FRET measurements by a beam splitter BS2 (HQ 575, AHF Tübingen). APD1 detected FRET donor fluorescence after an additional band pass F1 (BP 525/50, AHF Tübingen). APD2 detected the FRET acceptor photons after a long pass filter F2 (LP 595, AHF Tübingen).

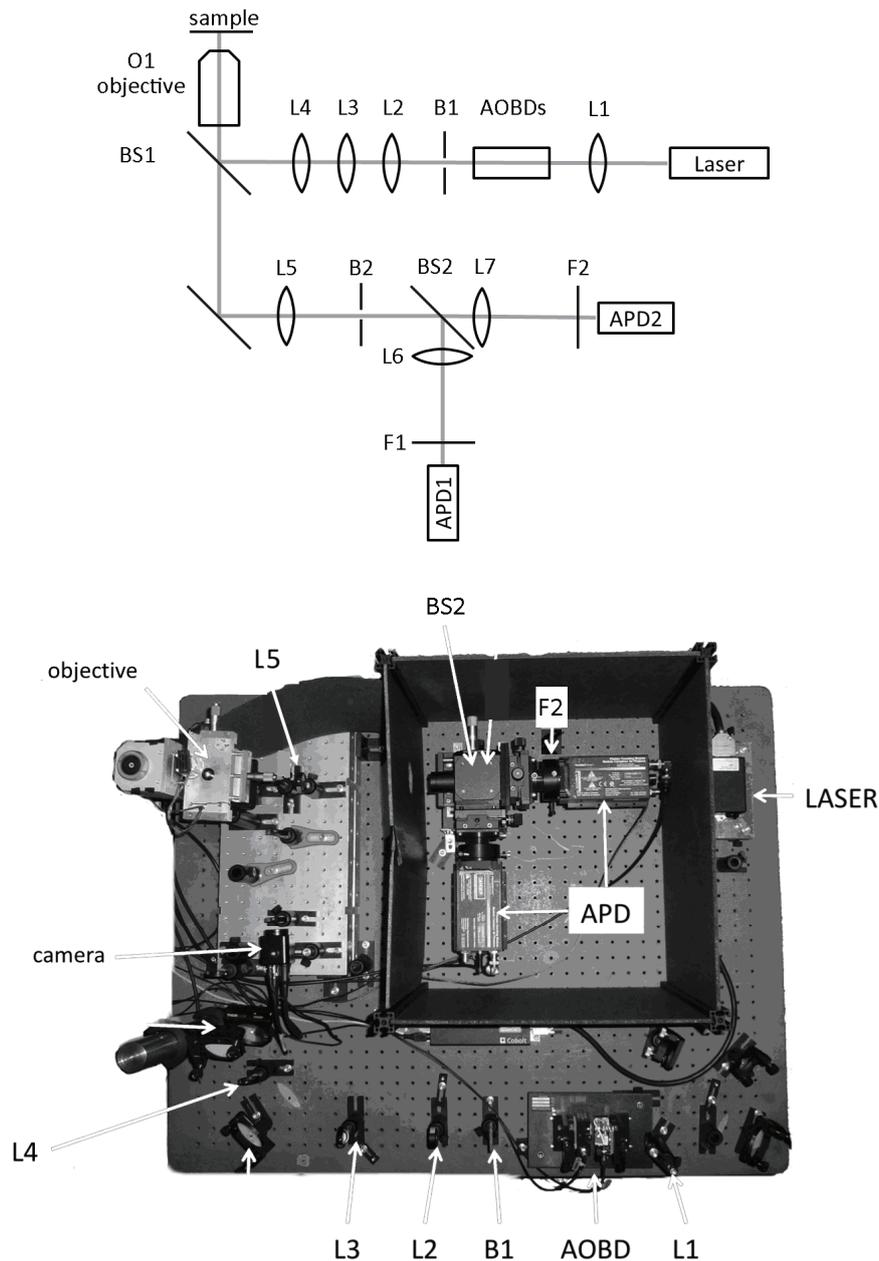

**Figure 1.** Scheme of the optical beam path of the ABELtrap and image of the microscope setup (see text for details).

The overall size of the ABELtrap setup is small and fits on an optical table with 70 x 90 cm dimensions (Figure 1b). The fluorescence detection path is enclosed by box of black cardboard to protect from ambient light. Both APDs were aligned and mounted together with the pinhole and the according optics onto a single x,y,z adjustable mechanical stage (OWIS Staufen, Germany). This facilitated the alignment procedures and resulted on a stable setup. An additional sensitive camera (WAT-120N+, WATEC) is integrated into the setup to enable inspection and calibration of the deflected laser beam.

### 2.2 Functions of the FPGA

The field-programmable gate array (FPGA) on a PCI card (FPGA 7813R, National Instruments) is the essential control component of the ABELtrap. The FPGA gets the TTL pulses (photon counts) from the APDs, and updates the voltages on the four electrodes in order to move the nanometer-sized particle to the trapping position within the sample chamber. Figure 2 shows the wiring of the different in- and outputs of the FPGA.

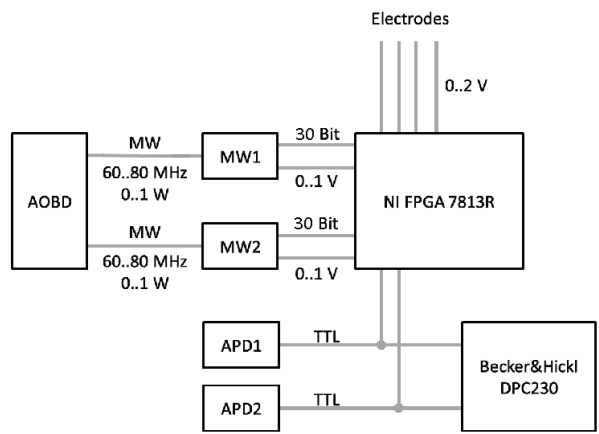

**Figure 2**. Scheme of the electrical wiring of inputs and outputs the FPGA, the AOBD, APDs and TCSPC card.

The spot of the laser beam can be moved in x and y directions by altering the microwave frequencies of the two laser beam deflectors (AOBDs). The AOBDs were operated by two digital microwave generators, which received the frequency values as 30 bit plus an additional analog value to set up the generated microwave power. With increasing power of the microwave, the intensity of the deflected laser beam increased in order to optimize the fluorescence signal on the APDs. Photon were recorded in parallel by a time-correlated single photon counting module (DPC-230 Becker&Hickl, Berlin, Germany) using a BNC splitter. Each photon was recorded with a time resolution of 165 ps. Photons could be recorded in up to eight independent channels simultaneously. The total measurement time was virtually unlimited. The software 'Burst Analyzer' (Becker&Hickl) was used for subsequent single-molecule data viewing and analysis, which is based on our previously developed program[7].

The main control unit of the setup, the FPGA, has to be programmed. We modified the FPGA software which was a public available LabView-code from A. E. Cohen and coworkers[40]. For generating the laser focus scan patterns, Cohen et al. used a set of two EOBDs which required an analog driver and expected analog voltages with 16 bit resolution to define the points of the laser scan pattern. For our different AOBDs we used a digital microwave synthesizer as the driver (Gooch & Housego, analog system digital frequency synthesizer 64020-200-2ADSDFS-A) which required 30 digital lines to set up the frequency values. Furthermore, one additional analog line is available to set and control the microwave power and, thereby, the laser power after the AOBDs.

The FPGA control software used an internal 32 bit memory block which represented the x,y-values of the laser scan pattern with 16 bit resolution (according to the resolution of the analog output units of the FPGA). One 32 bit value was set from a 16 bit x and a 16 bit y coordinate. The PC interface allowed the user to select one of several different laser scan patterns. Every time the user selects a new scan pattern, an updated list of x,y coordinates was transferred as 32 bit words to the FPGA to create the corresponding output voltage at its analog output unit.

The digital microwave synthesizer expected a 30 bit word at the input interface to define the microwave frequency for each AOBD. The AOBD has a middle frequency of 70 MHz for the first order of diffraction. The microwave synthesizer covered a frequency range of 50 to 90 MHz.

In order to maintain a high processing speed and to prevent unnecessary use of the limited onboard memory of the FPGA, we transferred only the difference of the actual frequency to the middle frequency as a 16 bit word for the x and y coordinates. The final 30 bit frequency values were composed by adding the constant value of the middle frequency to the 16 bit values stored in the intermediate buffer, and by shifting them before adding to exploit the full available frequency range.

### 2.3 Calibrating the AOBDs

A calibration factor of beam deflection in MHz to a laser spot displacement in µm in the focal plane had to be determined. Therefore we wrote a new FPGA software tool to image a scanned area. The AOBDs were driven to illuminate a rectangular field by the moving laser spot stepwise, and the FPGA transferred the number of collected photons per pixel together with the actual laser position to the software to generate the intensity image. Each pixel was false-colored representing the photon count rate. Thus the confocal set up could be used to create images of a limited area of about 2 x 2 μm, enough to image a single immobilized fluorescent molecule or nanocrystal.

Single fluorophores are point emitters and create a diffraction limited spot with a diameter $d$ defined by the wavelength of the fluorescent light λ according to[47]:

$$d = \frac{\lambda}{2 NA} \quad (1)$$

with NA, numerical aperture of the objective.

Our strategy of obtaining the calibration factor was to image a photostable fluorescent nanoparticle and to analyze the width of the diffraction limited spot in the image, given the theoretical diameter $d$ and an ideal point spread function of the microscope objective.

A single Nitrogen-Vacancy (NV) center within a surface-immobilized nanodiamond was well-suited for this calibration task. The emission wavelength of the luminescent NV center had a maximum at about 670 nm, which could be detected by our current setup using APD2. The NV center is extremely photostable, it does not show any blinking or bleaching when embedded deeply in the diamond[48]. Therefore, we obtained clean images at scan times in the range of minutes. The numerical aperture of the objective was 1.35 yielding a diameter $d$ = 248 nm.

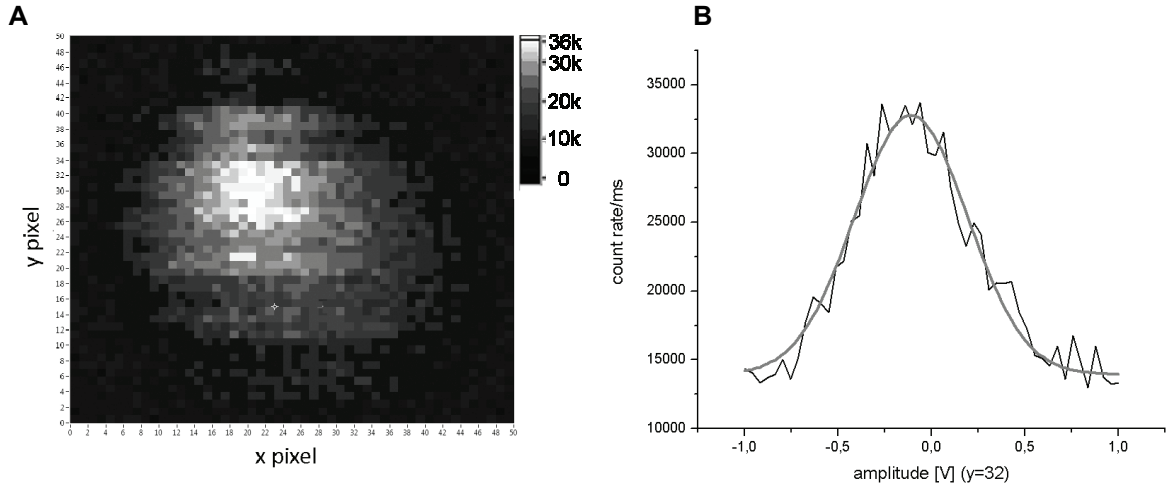

**Figure 3.** (**A**) Confocal image of a luminescent NV center of a surface-immobilized nanodiamond. The image size is 50 x 50 pixels, with the applied microwave voltages ranging from - 1 V to + 1 V in x and y direction. The photon count rate reached 36 counts per ms (36 kHz) at the brightest pixels in the middle of the image. (**B**) Gaussian fit of the imaged NV center. The width of the Gaussian fit corresponded to 0.30 V.

In the scan image shown in Figure 3A. A scan pattern of ±1 V divided into 50 points was generated. The Volt-values were transferred in frequencies by multiplying them with the constant factor 1/10 x 215 x 212=13.3 x 10$^6$ and adding these to the center frequency of 70 MHz. This yielded in a scan area from 56.6 MHz to 83.4 MHz for both AOBDs.

The Gaussian fit of one line (here in x direction) is shown in Fig. 3B. The width of the fit function is 0.30 V. The diameter corresponds to the full width at half maximum (FWHM) of a Gaussian function by:

$$d = FWHM = 2\sqrt{2\ln 2}\sigma \quad (2)$$

The expected diameter of 248 nm corresponded to a $\sigma$ of 105 nm. Comparing both values for the Gaussian width yielded the calibration or scaling factor of 0.35 μm/V.

## 2.4 Scan pattern

We used the given scan pattern in the LabView program of A. Fields and A. E. Cohen[40], i.e. a 19-point hexagonal pattern with point-to-point distances as shown in Figure 4 (below).

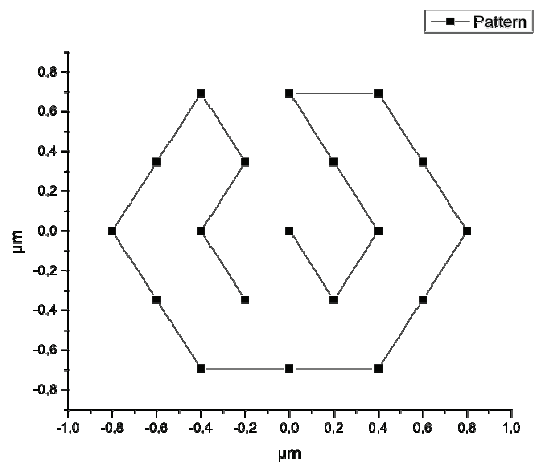

**Figure 4**. A 19 point hexagonal laser beam scan pattern was used in the ABELtrap.

The repetition rate of the laser beam scan pattern was set to 10 kHz, which corresponded to 5.26 µs per focus position. This turned out to be the maximum update rate for the AOBDs in order to achieve in a well-resolved position pattern. With higher update rates the points (positions) smeared out.

## 2.5 Additional operating parameters of the ABELtrap

All other operating parameters were used as stated in the publication of A. Fields and A. E. Cohen[40]. We limited the voltages applied to the platinum electrodes to ± 2 V to reduce electrochemical reactions at the electrodes.

The PDMS chamber design with a 1 µm shallow trapping region was described previously[46] and followed the design developed by A. E Cohen. For fabrication of the microfluidic trap chambers we used PDMS (Sylgard® 184 Elastomer Kit; Dow Corning) with a hardener in a petri dish.

## 3 RESULTS

To test the performance of the actual ABELtrap components and to evaluate the use for trapping soluble single proteins like the 10-nm small $F_1$ part of $F_oF_1$-ATP synthase in the future we used a diluted solution of fluorescent 20-nm beads (fluospheres 'yellow-green' 505/515, F8787, Molecular Probes). These nanobeads contain a high number of fluorophores ensuring a high signal-to-background ratio, which allowed us to reduce the laser excitation power and, therefore, to minimize the luminescence from the glass cover slip and PDMS sample chamber. Single fluorophore impurities in the solution did not influence the trapping of these bright nanobeads.

Figure 5 shows the time trace of a trapped nanobead and its escape from the ABELtrap. In the first 1600 ms the nanobead is trapped indicated by high photon count rates. The time resolution is 1 ms in Fig. 5A, that is averaged over 10 consecutive 19-points patterns, or 190 consecutive focus points to one time bin, respectively. However, the nanobead was not trapped stationary, but could escape several times, and was re-trapped. This trapping behavior could be analyzed to identify performance deficits of our setup. In Fig. 5B, the time trace of the nanobead position shows the actual deviation from the center of the trap. This demonstrates that the nanobead was not pushed back to the center at distance '0 nm' but was trapped at a position about 150 nm off the center position. The time trajectory of the applied voltages at the platinum electrodes in Fig. 5C revealed that during trapping of the bead, the voltages are below 100 mV, and increased to 70 (±30 mV) after escape from the trap. At these low potentials we do not expect electrochemical reactions at the electrodes. The one-dimensional distribution of the position deviations of the trapped nanobead in Fig 5D had one maximum at 105 nm and a FWHM around 200 nm, whereas the apparent deviation distribution of 'background noise' after 1600 ms was broadened with two distinct maxima.

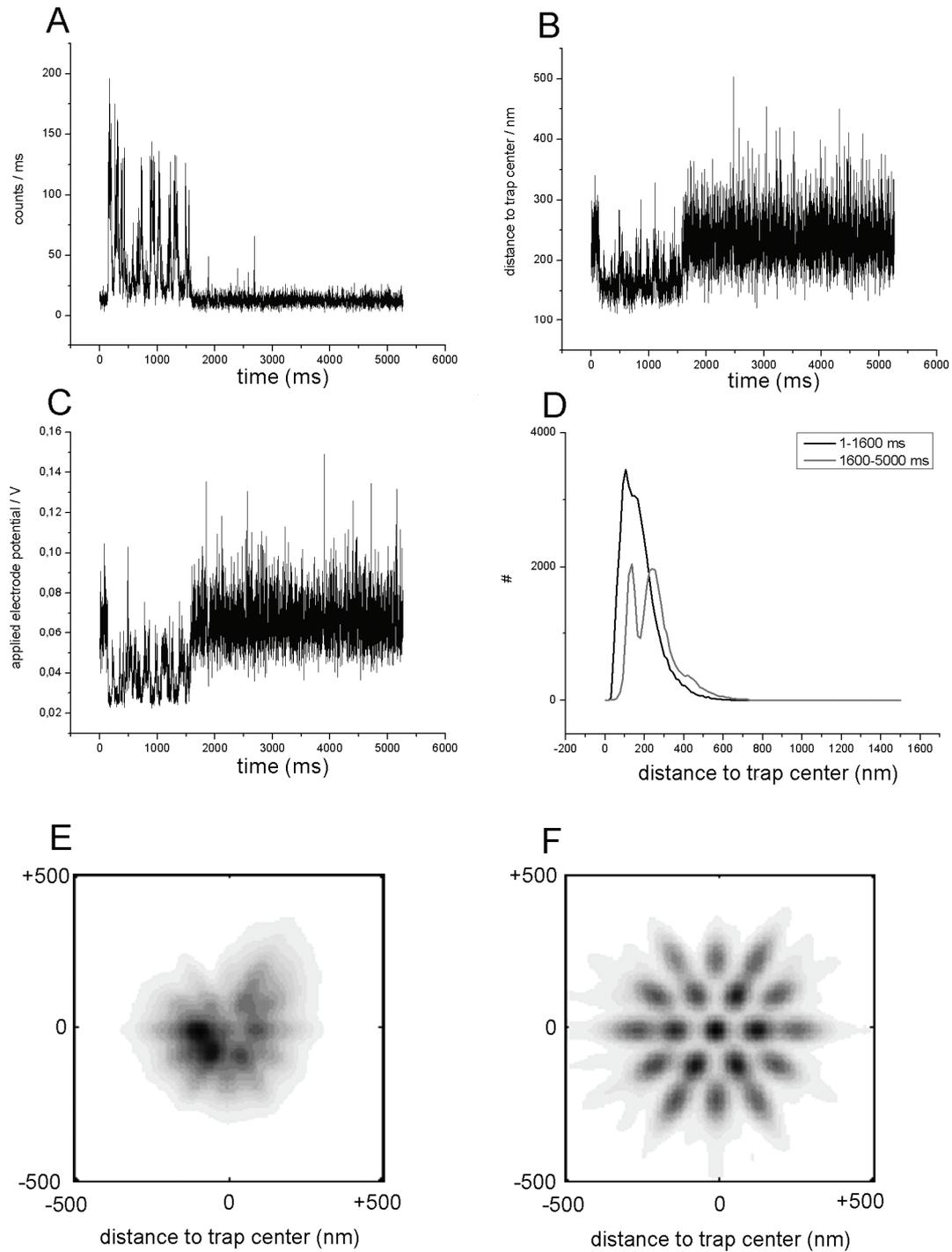

**Figure 5.** Time trace analysis of a trapped fluorescent polystyrene nanobead. The applied laser beam pattern was a 19 point hexagonal with a frame update rate of 10 kHz. (**A**) Intensity trajectory during trapping of a bead. The time resolution is 1 ms. In the first 1600 ms the nanobead was trapped but escaped several times and was re-trapped. (**B**) Corresponding time trace of the bead position deviation from the trap center. (**C**) Corresponding time trace of the applied voltages to the electrodes. (**D**) Histograms of the mean distances to the trapping center. The first 1600 ms with the trapped nanobead exhibit a most frequent distance of 105 nm away from the center, whereas in the absence of a trapped bead the two position maxima matched exactly the two rings of the hexagonal pattern at 135 nm and 240 nm. (**E**) Histogram of the estimated nanobead positions during trapping. The most frequent position of the nanobead was shifted asymmetrically from the center of the scan pattern in -x and -y direction. (**F**) Distribution of position probabilities in the ideal symmetrical case for the 19-points hexagonal focus pattern.

The time traces of the position deviation of this trapped nanobead as well as the apparent deviation after escape from the trap were analyzed further by separating the x and y directions (Fig. 5E). The non-symmetric distribution of nanobead positions clearly indicated ABELtrap performance deficits, either owing to non-uniform illumination of the focus pattern, or due to non-symmetric shapes of the PDMS chamber. In addition, the actual potentials on each of the four electrodes, as set by the FPGA, might not be identical. In the ideal case of 19 non-overlapping focus positions of the laser pattern, a symmetric distribution with an inner and an outer ring should be obtain (Fig. 5F). The apparent deviation of the background from the center of the trap showed exactly these two ring-like distances in Fig. 5D.

Using another PDMS chamber and optimized trapping parameters for the nanobeads, we succeed in trapping of these 20-nm fluorescent polystyrene beads for up to 8 seconds (Fig. 6). Slow photobleaching revealed that the nanobead contained multiple fluorophores. Intensity oscillations during trapping unraveled remaining performance deficits of our setup. However, the possibility to trap a single small particles with this version of an ABELtrap was demonstrated.

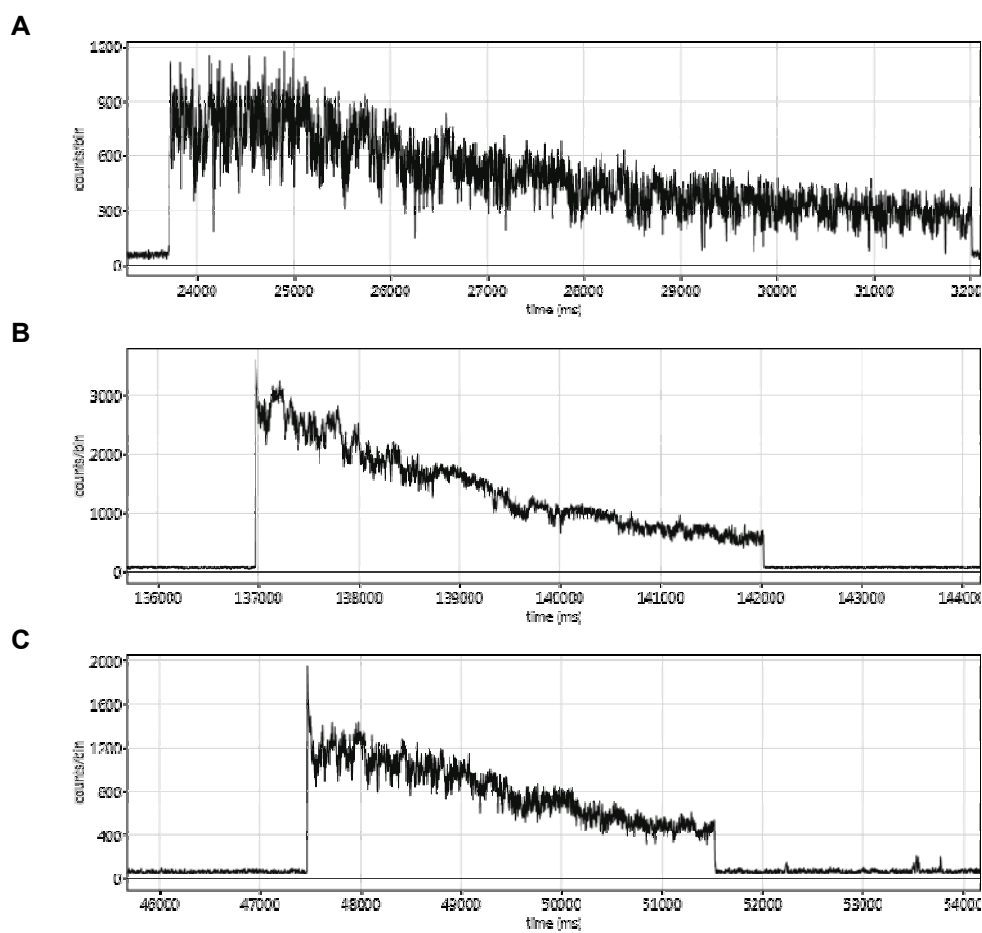

**Figure 6.** Fluorescence time trajectories of three different 20-nm nanobeads in the ABELtrap. (**A**) Individual trapping dwells were achieved for more than 8 seconds, (**B**) for about 5 seconds, and (**C**) for about 4 seconds. The time axis with 1 ms time resolution shows elapsed time after starting the measurements.

## 4 DISCUSSION

Since the first appearance and publications of the design und function of an ABELtrap in 2005, this single-molecule manipulation device has been developed rapidly. Latest versions achieve a confinement of individual fluorescent molecules in solution at the one nanometer limit[40]. Given the benefits of holding a single molecule in solution and thereby avoiding surface-attachment problems, implementations of ABELtraps at various research laboratories seem to be desirable.

The requirements to set up an ABELtrap are not only optical components, but also the control software and the trapping chambers. Nanofabrication utilities are necessary to make the silicon wafers for PDMS cells. The quartz cells are significantly more demanding in the etching processes. They are expensive and are not considered to be consumables. Therefore, the possibility of using disposable PDMS cells is the cheaper alternative and easier to handle, if it is possible to trap the molecule of interest because the drawback of PDMS-cover glass cells is the higher background photon count rate due to luminescent impurities.

With the publicly available software from A. E. Cohen group (Harvard) to run a fast FPGA-based ABELtrap, a new implementation of an ABELtrap was enormously facilitated. We needed to re-program only parts of the code to control our AOBDs instead of the EOBDs in the original version. Thus, we expect that more single-molecule microscopy groups will set up these types of ABELtraps and might contribute to solve some important issues. Surface aestivation methods to prevent sticking of the biopolymers have to be optimized. The effects of the variable electric fields on enzymes and their conformational dynamics have to be determined, better photostable and bright fluorophores are required, and ways to control the luminescent background in the PDMS-glass chambers have to be found.

Our interest is to use an ABELtrap for soluble enzymes and liposome-reconstituted membrane transporters that might work slowly, for example the $F_oF_1$-ATP synthases at low [ATP] or at low proton motive force, but also the ATP-driven $K^+$ transporter KdpFABC[49] (a P-type ATPase), the ABC transporter Pgp[50], as well as vacuolar proton pumps like V-ATPases[51, 52]. We want to be able to record time trajectories of conformational changes also for other soluble ATP-driven motors like SecA (part of the protein translocation machinery, but not the YidC pathway[53]) or the soluble $F_1$ fragments[1]. First intermediary objectives towards these aims have been achieved with the demonstration of trapping 20-nm fluorescent polystyrene beads in solution for up to 8 seconds in our version of an ABELtrap.

**Acknowledgements**

Financial support by the Baden-Württemberg Stiftung (by contract research project P-LS-Meth/6 in the program "Methods for Life Sciences") is gratefully acknowledged. This work was supported in part by the DFG grant BO 1891/16-1 to M.B.. The authors want to thank Prof. Dr. S. D. Dunn (Western University of Ontario, London, Canada), Prof. Dr. T. Duncan (SUNY Upstate Medical University, Syracuse, NY, USA) and Prof. Dr. P. Gräber (University of Freiburg, Germany) for their ongoing support in $F_oF1$-ATP synthase mutants and enzyme preparations, and Prof. Dr. J. Wrachtrup (Stuttgart University, Germany) for supply of fluorescent nano-diamonds.